\input harvmac
\input epsf
\Title{\vbox{\baselineskip12pt\hbox{UGVA-DPT 1998/02-998}
\hbox{DFUPG 1-98}}}
{\vbox{\centerline{Geometric Aspects of }
\vskip2pt\centerline{Confining Strings}}}
\centerline{M. Cristina Diamantini}
\centerline{Dipartimento di Fisica, Universit\`a di Perugia}
\centerline{via A. Pascoli, I-06100 Perugia, Italy}
\bigskip
\centerline{Carlo A. Trugenberger}
\centerline{D\'epartement de Physique Th\'eorique, Universit\'e de Gen\`eve}
\centerline{24, quai E. Ansermet, CH-1211 Gen\`eve 4, Switzerland}

\vskip .3in
\noindent
Confining strings in $4D$ are effective, thick strings 
describing the confinement
phase of compact $U(1)$ and, possibly, also 
non-Abelian gauge fields. We show that
these strings are dual to the gauge fields, inasmuch their perturbative regime
corresponds to the strong coupling ($e\gg 1$) regime of the gauge theory. In
this regime they describe smooth surfaces with long-range correlations and
Hausdorff dimension two. For lower couplings $e$ and monopole fugacities $z$,
a phase transition takes place, beyond which the smooth string picture is lost.
On the critical line intrinsic distances on the surface diverge and
correlators vanish, indicating that world-sheets become fractal.

%\draft
\Date{March 1998}

\lref\gsw{For a review see e.g.: 
M. B. Green, J. H. Schwarz and E. Witten, ``{\it Superstring 
Theory}", vol. 1, Cambridge University Press, Cambridge (1987).} 

\lref\polbook{For a review see e.g.: 
A. M. Polyakov, ``{\it Gauge Fields and Strings}", Harwood Academic
Publishers, Chur (1987).}

\lref\polrev{A. M. Polyakov, {\it Physica Scripta} {\bf T15} (1987) 191.}

\lref\pola{A. M. Polyakov, {\it Nucl. Phys.} {\bf B268} (1986) 406.}

\lref\polb{A. M. Polyakov, {\it Nucl. Phys.} {\bf B486} (1997) 23.}

\lref\ka{H. Kleinert, {\it Phys. Lett.} {\bf B174} (1986) 335.}

\lref\kb{H. Kleinert and A. Chervyakov, ``{\it Evidence for
Negative Stiffness of QCD Strings}", hep-th/9601030.}

\lref\larged{H. Kleinert, {\it Phys. Rev. Lett.} {\bf 58} (1987) 1915;
P. Olesen and S. K. Yang, {\it Nucl. Phys.} {\bf B283} (1987) 73;
E. Braaten, R. D. Pisarski and S. M. Tze, {\it Phys. Rev. Lett.} 
{\bf 58} (1987).}

\lref\david{F. David and E. Guitter {\it Nucl. Phys.} {\bf B295}
(1988) 332, {Europhys. Lett. } {\bf 3} (1987) 1169.}

\lref\bz{E. Braaten and C. K. Zachos, {\it Phys. Rev. } {\bf D35} (1987) 1512.}

\lref\qt{F. Quevedo and C. A. Trugenberger, {\it Nucl. Phys.} {\bf B501}
(1997) 143.}

\lref\dqt{M. C. Diamantini, F. Quevedo and C. A. Trugenberger, {\it Phys. Lett.}
{\bf B396} (1997) 115.}

\lref\dt{M. C. Diamantini and C. A. Trugenberger, ``{\it Surfaces with
Long-Range Correlations from Non-Critical Strings}", hep-th/9712008,
to appear in {\it Physics Letters} {\bf B}.}

\lref\dtfut{M. C. Diamantini and C. A. Trugenberger, in preparation}

\lref\kr{V. I.
Ogievetsky and V. I. Polubarinov, {\it Sov. J. Nucl. Phys.} {\bf 4}
(1967) 156; M. Kalb
and P. Ramond, {\it Phys. Rev.} {\bf D9} (1974) 2273.}

\lref\orla{P.Orland, {\it Nucl. Phys.} {\bf B428} (1994) 221.}

\lref\gr{I. Gradstheyn and I. M. Ryzhik, ``{\it Table of Integrals, Series
and Products}", Academic Press, Boston (1980).}

\lref\wi{E. Witten, {\it Phys. Lett. } {\bf B86} (1979) 283.}

\lref\polcha{For a review see e.g. :
J. Polchinski, ``{\it Strings and QCD}",
contribution in Symposium on Black Holes, Wormholes
Membranes and Superstrings, H.A.R.C., Houston (1992); hep-th/9210045.}

\lref\ps{J. Polchinski and A. Strominger, {\it Phys. Rev. Lett.} {\bf 67}
(1991) 1681.}

\lref\pz{J. Polchinski and Z. Yang, {\it Phys. Rev.} {\bf D46} (1992) 3667.}

\lref\cara{J. C. Cardy and E. Rabinovici, {\it Nucl. Phys.} {\bf B205} (1986) 1.} 

\lref\dfj{B. Durhuus, J. Fr\"ohlich and T. Jonsson, {\it Nucl. Phys.}
{\bf B240} (1984) 453; B. Durhuus and T. Jonsson, {\it Phys. Lett.}
{\bf 180B} (1986) 385.}

\lref\zinn{For a review see e.g.: J. Zinn-Justin, ``{\it Quantum Field
Theory and Critical Phenomena}", Oxford University Press, Oxford (1996).}

\newsec{Introduction}
It is an old idea that the confining phase of gauge theories can be
formulated as a string theory \polcha . The natural relevant term in the
action of such a string is the Nambu-Goto term, proportional to the area
of the world-sheet. However, this term does not lead to a consistent
theory outside the critical dimension 26 \gsw . 

The Nambu-Goto term describes fundamental strings, which do not have a
transverse extension. On the other hand, one can convince oneself on 
general grounds \ps \ that the strings describing electric flux tubes
in QCD must be thick strings, with a fundamental transverse scale and therefore
the theory of these objects is an {\it effective string theory} .

In order to take into account the bending rigidity due to the finite width
and to cure the problems of the fundamental 
Nambu-Goto action, Polyakov \pola \ and
Kleinert \ka \ proposed to add to it a marginal term proportional to
the extrinsic curvature of the world-sheet.
The so obtained {\it rigid string}, however, is plagued by various problems
of both geometric and physical nature. From the geometric point of
view, the new term turns out to be infrared irrelevant \refs{\pola ,\ka}
\ and violent fluctuations lead to the formation of a finite correlation
length for the normals to the surface and to crumpling \refs{\larged ,\david} ,
which is unacceptable for QCD strings. 
From the physical point of view, the new term brings about an unphysical
ghost pole in the propagator \ka \ and the spectrum is non-unitary and
unbounded by below \refs{\bz , \polcha} .
Moreover, the high-temperature free energy of rigid strings has the
opposite sign to the result obtained from large-$N$ QCD \pz , although
the $\beta $-dependence comes out correct.

Recently, Polyakov \polb \ (see also \qt ) proposed a new action to describe
the confining phase of gauge theories. This {\it confining string} theory
can be explicitly derived \dqt \ for a $4D$ compact $U(1)$ gauge theory
in the phase with a condensate of magnetic monopoles \polbook . Polyakov
\polb \ conjectured moreover that the only modification for non-Abelian
gauge fields should be the inclusion in the string action of a corresponding
group factor. In $4D$, the confining string is indeed an effective string
with a microscopic length scale describing the thickness of the string.
The major differences with respect to the rigid string are a non-local
interaction between world-sheet elements and a {\it negative stiffness} \dqt .

A different, but essentially equivalent formulation of the confining string
was considered by Kleinert and Chervyakov \kb , who showed that the
high-temperature free energy matches the large-$N$ QCD result also in sign,
and that no unphysical ghost pole is present.
In \dt \ we showed that world-sheets of confining strings are characterized
by {\it long-range correlations} for the normals, due to a non-local
``antiferromagnetic" interaction. Moreover, it is easy to convince oneself
that the spectrum of confining strings is perfectly 
bounded by below \dtfut .
So, confining strings are very promising, given that they seem to solve all
the problems of rigid strings.

In this paper we address the quantum phase structure and geometric aspects
of confining strings. After a review of the confining string action in
section 2, in section 3 we compute the one-loop correction $t_1$ to the
classical string tension $t_0$ in the semiclassical expansion. This
allows us to find the combinations of the two dimensionless parameters of
the theory, the gauge coupling $e$ and the monopole fugacity $z$, for which
$|t_1/t_0| \ll 1$. We show that this perturbative regime corresponds to large
coupling $e$, thereby proving the {\it duality} between confining strings
and gauge fields. By computing correlators for the normals and estimating
the Hausdorff dimension we show that, in the perturbative regime, the
world-surfaces are {\it smooth} objects with intrinsic dimension two.

In section 4 we describe the modifications induced by the presence of a
$\theta $-term in the gauge theory; notably we show how duality is modified 
in this case. 

In section 5 we use the formulation of Kleinert and Chervyakov to compute
the quantum phase structure of the theory in the large-$D$ expansion.
We show that smooth strings exist only below a {\it critical line} in the
$(z, 1/e)$ plane: this domain matches the perturbative region
found with the one-loop calculation in section 3. The critical line
is characterized by the divergence of intrinsic distances on the surface
and the vanishing of normals correlators. 

Finally, we draw our conclusions in section 6. 
 
\newsec{Confining strings}
Confining strings have an action which is induced by a Kalb-Ramond 
antisymmetric tensor field \kr . In $4$-dimensional Euclidean space 
it is given by
\eqn\csa{\eqalign{{\rm exp} \left(-S_{CS} \right) &= {G\over 
Z\left( B_{\mu \nu} \right)} 
\ \int {\cal D}B_{\mu \nu} \ {\rm exp} \left\{ -S\left( B_{\mu \nu } \right)
+i \int d^4 x\ B_{\mu \nu }T_{\mu \nu} \right\} \ ,\cr
T_{\mu \nu} &= {1\over 2} \ \int d^2{\sigma }\ X_{\mu \nu} ({\sigma})
\ \delta ^4 ({\bf x}-{\bf x({\sigma})}) \ ,\cr
X_{\mu \nu} &= \epsilon^{ab} {\partial x_{\mu }\over \partial {\sigma }^a}
{\partial x_{\nu}\over \partial {\sigma }^b} \ ,\cr }}
with ${\bf x}({\sigma })$ parametrizing the world-sheet and $G$ the group
factor characterizing the underlying gauge group.
Given that $G$ is of no importance for the following, we will henceforth
set $G$ to its value $G=1$, valid for a compact $U(1)$ group.
At long distances
the action for the Kalb-Ramond field reduces to
\eqn\akrf{\eqalign{S\left( B_{\mu \nu } \right) &= \int d^4 x \ {1\over 12
z^2 \Lambda ^{2}}\ H_{\mu \nu \alpha}H_{\mu \nu \alpha} + {1\over 4e^2}
B_{\mu \nu }B_{\mu \nu }\ ,\cr
H_{\mu \nu \alpha } &\equiv \partial_{\mu }B_{\nu \alpha}+\partial_{\nu}
B_{\alpha \mu}+\partial_{\alpha}B_{\mu \nu} \ .\cr}}
It depends on a short-distance cutoff $\Lambda $ and
on two dimensionless parameters $e$ and $z$.
This action can be explicitly derived \refs{\polb ,\dqt } \ from 
a lattice formulation of compact QED in the phase
with condensing magnetic monopoles \polbook , and constitutes a special
case of a generic mechanism of $p$-brane confinement proposed in \qt .

In the lattice model $1/{\Lambda}$ plays the role of the lattice
spacing while $z^2$ is the monopole fugacity; $e$ is the coupling constant
of the underlying gauge theory. Note that, for $z^2\to 0$, only configurations
with $H_{\mu \nu \alpha}=0$ contribute to the partition function: this means
that the Kalb-Ramond field becomes pure gauge, $B_{\mu \nu} = \partial_{\mu}
A_{\nu}-\partial_{\nu}A_{\mu}$, and we recover the partition function of
QED coupled to point-particles described by the boundaries of the original
world-sheet. 

In the continuum formulation above, $\Lambda $ can be viewed
as a Higgs mass and, correspondingly, $1/\Lambda$ as a finite thickness of
the string. The mass of the Kalb-Ramond 
field is given by 
\eqn\mass{m=\Lambda {z\over e}\ .}
The dimensionless parameter 
\eqn\type{\tau \equiv {m \over \Lambda} = {z\over e}\ ,}
plays thus the same role as the ratio (coherence length/penetration
depth) in superconductivity
theory. This close analogy with superconductivity is not surprising
when one realizes that the same action has been derived for magnetic vortices
in the framework of the Abelian Higgs 
model \orla .

In \dt \ we have shown that, up to boundary terms (which are of no consequence
in the present paper), the confining string action can be rewritten as
\eqn\ist{S_{CS}=\Lambda ^2 \ \int d^2\sigma \sqrt{g} \ \ t_{\mu \nu }
(\sigma ) \ G\left( z, e, 
\left( {{\cal D}\over \Lambda}\right) ^2\right) \ t_{\mu \nu }
(\sigma ) \ ,}
where we have introduced the induced metric 
\eqn\inme{\eqalign{g_{ab} &\equiv \partial_a x_{\mu } \partial_b x_{\mu}\ ,\cr
g &\equiv {\rm det}\ g_{ab}\ ,\cr}}
the tangent tensor
\eqn\tate{t_{\mu \nu } \equiv {1\over \sqrt{2g}}\ X_{\mu \nu }\ ,}
and the covariant Laplacian
\eqn\cola{{\cal D}^2 = {1\over \sqrt{g}} \partial _a g^{ab} \sqrt{g}
\partial _b \ .}
The Green's function $G$ is defined as the Taylor series obtained from
the generating function 
\eqn\sposs{G\left( z, e, 
\left( {{\cal D}\over \Lambda}\right) ^2\right)=
{z^2\over {4 \pi}}\ K_0 \left( \sqrt{\tau ^2 
- \left( {{\cal D}\over \Lambda}
\right)^2 } \right) \ ,}
with $\tau $ defined in \type , and $K$ a modified Bessel function \gr .
Its first few terms are given by
\eqn\expa{\eqalign{G &\left( z, e, 
\left( {{\cal D}\over \Lambda} \right) ^2 \right)
= t_0 + s \left( {{\cal D}\over \Lambda } \right) ^2 + 
w \left( {{\cal D}\over \Lambda } \right) ^4 +\dots \ ,\cr
t_0 &= {z^2\over {4 \pi }} \ K_0 (\tau ) \ ,\cr
s &={z^2\over {8 \pi }}\ \tau ^{-1} \ K_1 (\tau ) \ ,\cr
w &={z^2\over {32 \pi }}\ \tau ^{-2} \ K_2 (\tau )
\ .\cr }}
When compared with the corresponding kernel of rigid strings \refs{\pola ,
\ka } ,
\eqn\rigid{G^{\rm rigid} = {\mu _0 \over \Lambda ^2} - {1\over \alpha}
\left( {{\cal D}\over \Lambda } \right) ^2 \ ,}
(with $\mu _0$ the bare string tension) this expression exposes 
best the two crucial aspects of confining strings: a {\it non-local}
interaction between surface elements and a {\it negative stiffness} $s$.
Contrary to the case of rigid strings, where the ``local ferromagnetic" 
interaction \rigid \ is not sufficient to prevent crumpling \refs{\larged ,
\david } \ the ``non-local antiferromagnetic" interaction \expa \ does indeed
lead to smooth strings \dt .

\newsec{Perturbative saddle point analysis}
From \ist , \sposs  \ and \expa \ one would naively conclude that for 
$e \to 0$ one can remove the cutoff $\Lambda $, so that $e=0$ is an
infrared fixed point corresponding to the usual Nambu-Goto string. In the
following we show that this is wrong, since the perturbative smooth regime
at strong coupling is separated from the naive Nambu-Goto regime $e\to 0$
by a phase transition.

To this end we shall use standard saddle-point techniques to investigate
the role of transverse fluctuations $\chi ^i (\sigma )$ around a long, straight
string configuration parametrized in the Gauss map as
\eqn\gauss{x_{\mu } (\sigma ) = \left( \sigma _0, \sigma _1, \chi ^i (\sigma )
\right) \ ,\qquad \qquad i=2, 3 \ ,}
where $-\beta /2\le \sigma_0 \le \beta /2$, $-R/2 \le \sigma ^1 \le R/2$.

This analysis is very simple for the confining string action. We start by
integrating out the Kalb-Ramond field, which appears quadratically in the
induced action \csa . Up to (here irrelevant) boundary terms we obtain
\eqn\nlf{S_{CS} = {z^2\Lambda ^{2}\over 4} \int d^2{\sigma} \int d^2
{\sigma }' \ X_{\mu \nu}({\sigma}) \ Y\left( {\bf x}
({\sigma }) - {\bf x}({\sigma}') \right) \ X_{\mu \nu}({\sigma }') \ ,}
where $Y$ is the $4$-dimensional Yukawa Green's function 
\eqn\dygf{Y(|{\bf x}|)= {m^2\over {4 \pi ^2}}
\ {1\over (m|{\bf x}|)} \ K_1 (m|{\bf x}|)\ ,}
and $K$ is a modified Bessel function \gr .
In order to regulate ultraviolet divergences we shall substitute this
Green's function with
\eqn\regf{Y' (|{\bf x}|) = {m^2\over {4\pi ^2}}
\ {1\over {m\sqrt{|{\bf x}|^2+{1\over 
\Lambda ^2}}}} \ K_1 \left( m 
\sqrt{|{\bf x}|^2 + {1\over \Lambda ^2}} \right) \ ,}
so that the potential is cut off on the scale $1/\Lambda $ corresponding to
the string thickness. Moreover, for simplicity, we shall use henceforth
dimensionless variables by measuring all distances in units of 
the thickness $1/\Lambda $
and all momenta in units of $\Lambda $:
\eqn\dim{\eqalign{\xi ^a &\equiv \Lambda \sigma ^a \ ,\cr 
{\bf r}_{\mu } &\equiv \Lambda {\bf x}_{\mu } \ ,\cr
r &\equiv \Lambda |{\bf x}|  \ ,\cr
\phi ^i &\equiv \Lambda \chi ^i\ .\cr }} 

In the Gauss map \gauss \ the components of the tensor $X_{\mu \nu }$ in
\csa \ take the following form:
\eqn\tensor{\eqalign{X_{01} &= - X_{10} = 1 \ ,\cr
X_{0i} &= -X_{i0} = {\partial \phi ^i \over \partial \xi ^1} \ ,\cr
X_{i1} &= -X_{1i} = {\partial \phi ^i \over \partial \xi ^0} \ ,\cr
X_{ij} &= -X_{ji} = O\left( (\phi )^2 \right) \ .\cr}}
Therefore, retaining only quadratic terms in the action, we are lead to
\eqn\saddle{S_{CS} = S_0 + S_1\ ,}
where $S_0$ represents the classical part, which, in the infinite area
($A=\beta R$) limit, is given by 
\eqn\cla{S_0 = t_0 \ \Lambda ^2 A\ .}
Here $t_0$, given in \expa , is the classical contribution to the string
tension (in dimensionless units). The contribution from transverse 
fluctuations is
\eqn\flu{\eqalign{S_1 &= \int d^2 \xi \int d^2 \xi ' \ \phi ^i (\xi )
\ V(\xi -\xi ') \ \phi ^i (\xi ') \ ,\cr
V(\xi ) &= V_1(\xi ) + V_2(\xi ) - \delta ^2(\xi ) \int d^2 \xi ' \ V_2(
\xi ') \ ,\cr
V_1(\xi ) &= \left( -\nabla ^2 \right) \ {z^2\tau ^2 \over 8\pi ^2} \ {1\over 
{\tau \sqrt{|\xi |^2 +1}}} \ K_1\left( \tau \sqrt{|\xi |^2+1} \right) \ ,\cr
V_2(\xi ) &= {z^2\tau ^2 \over 16 \pi ^2} \ {1\over {|\xi |^2 +1}} \ K_2 \left(
\tau \sqrt{|\xi |^2+1} \right) \ .\cr }}
The term $V_1$ arises from keeping linear terms in $\phi ^i$ in $X_{\mu \nu }$
and setting $\phi ^i=0$ in the kernel $Y$. The second term $V_2$, instead,
originates from an expansion to second order in $\phi ^i$ of the kernel $Y$
in \nlf \ while keeping only the zeroth-order of $X_{\mu \nu }$. The
$\delta $-function subtraction from $V_2$ takes into account that 
$Y\left( {\bf r} (\xi )- {\bf r} (\xi ') \right)$ in \nlf \ depends only
on differences $\left( \phi ^i (\xi ) - \phi ^i (\xi ')\right) $.

At this point we integrate over the two transverse fluctuations to obtain the
effective action
\eqn\eff{S_{CS}^{\rm eff} = t\ \Lambda^2 A\ ,}
with
\eqn\rst{\eqalign{t &= t_0+t_1\ ,\cr
t_1 &= {1\over 2\pi }\ \int_0^1 dp \ p \ {\rm ln} V(p)\ ,\cr }}
the renormalized string tension (in dimensionless units). Note that $p$
is the momentum in units of the short-distance cutoff $\Lambda $: this is
why the integral is cut off at one. The Fourier transform 
\eqn\fourier{V(p)= \int d^2\xi \ V(\xi ) \ {\rm e}^{ip\xi }\ ,}
of the fluctuation kernel \flu \ can be computed analytically:
\eqn\fouflu{\eqalign{V(p) &= V_1(p) + V_2(p)\ ,\cr
V_1(p) &= {z^2\over 4\pi} \ p^2 K_0\left( \sqrt{\tau ^2 + p^2} \right) \ ,\cr
V_2(p) &= {z^2\over 8\pi } \ \left\{ \sqrt{\tau ^2 + p^2}
\ K_1\left( \sqrt{\tau ^2
+p^2} \right) - \tau \ K_1(\tau ) \right\} \ .\cr }}
Note that $V(p)\propto p^2$ for $p\ll 1$; this shows that, at large
distances, the transverse fluctuations are described by 2 free bosons,
as expected.

Naturally, the expression \rst \ for the renormalized string tension
makes sense only in a perturbative region of parameter space where the ratio
\eqn\ratio{r(z, e) \equiv | {t_1(z, e)\over t_0(z,e)} | \ll 1\ ,}
even more so, given that $t_1$ is negative for most parameters and it is
easy to get a negative overall string tension. To give an idea of the
perturbative region of parameter space we choose an arbitrary cutoff at
20 \% and we plot in fig. 1 the region $r<0.2$ as a function of $z$ and $1/e$.

\vskip 30 pt                        
                         
\epsfbox{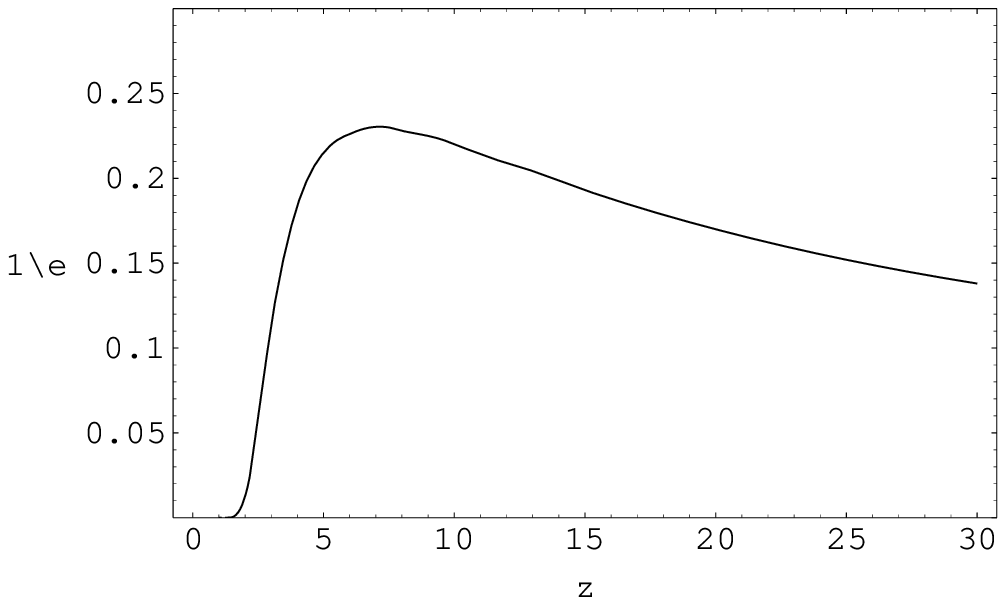}

\vskip 3pt
{\centerline {Fig. 1: the curve defining the upper boundary of the
perturbative region.}}
\vskip 30pt

\noindent
We see that the perturbative regime is characterized by large couplings $e$
and large monopole fugacities $z$. This is a first important result:
confining strings are indeed {\it dual} to compact $U(1)$ gauge fields in the
sense that the perturbative regime for the string corresponds to the strong
coupling regime for the gauge theory. Moreover, one cannot take the limit
$e\to 0$ to remove the cutoff and obtain the Nambu-Goto string. In doing so
the renormalized string tension decreases, as shown in fig. 2, until one reaches
the non-perturbative region where wild transverse fluctuations destroy the
string and a phase transition takes place (see section 5). 

\vskip 30 pt                        
                         
\epsfbox{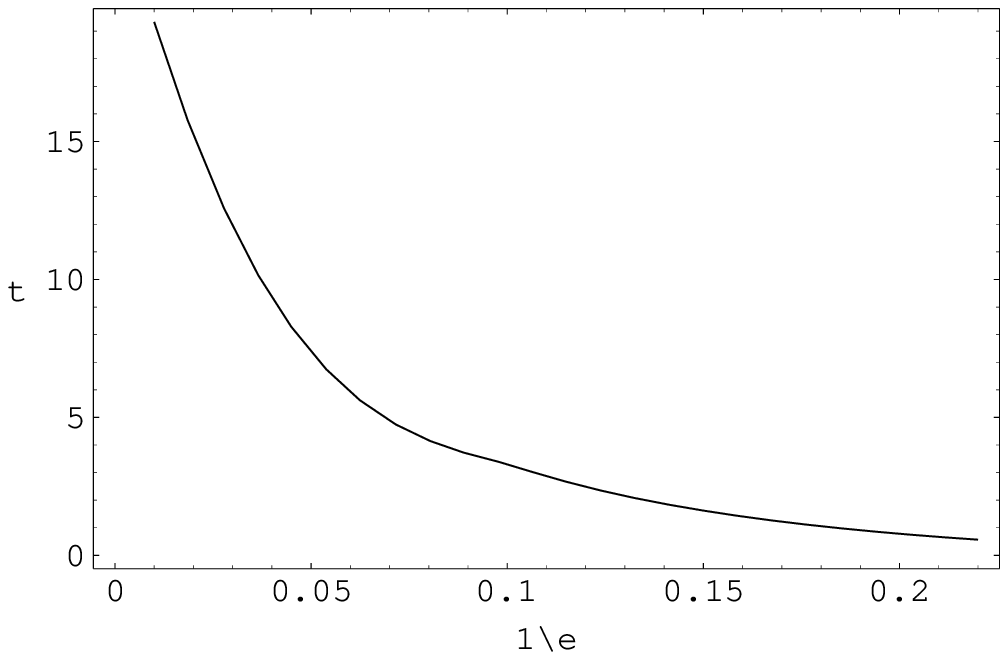}

\vskip 3pt
{\centerline  {Fig. 2: the string tension $t$ (dimensionless units)
for $z=10$.}}
\vskip 30pt

In the following we are going to investigate some geometric properties of
the confining string in the perturbative region derived above. First
of all we compute the {\it correlation function}
\eqn\corr{g_{ab}(\xi -\xi ') \equiv \langle \partial_a \phi ^i (\xi )
\ \partial_b \phi ^i (\xi ') \rangle \ ,}
for the scalar product of the components of the tangent vectors
normal to the reference plane $\left( \xi^0, \xi ^1 \right)$ at different
points on the surface. This correlation provides a picture of the role
of transverse fluctuations as a function of the parameters $z$ and $e$. It is
obtained form \flu \ and \fouflu \ as
\eqn\corrb{g_{ab} \left( \xi - \xi ' \right) = \delta _{ab} 
\ {1\over (2\pi )^2}\ \int d^2p \ {p^2\over 2V(p)}
\ \ {\rm e}^{i p (\xi -\xi ')} \ .}
The requirement on $g_{ab}(\xi )$ is that its inverse Fourier transform
reproduces the correct behaviour
\eqn\ex{{V(p)\over p^2} = {3\over 4} t_0 - {7\over 8} s \ p^2 + O(p^4)\ ,}
with $t_0$ and $s$ given in \expa , in the region of small $p$,
where the fluctuations reduce to free bosons.

Using only the expansion \ex \ the correlation
function \corr \ can be computed
independently of the ultraviolet details of $V(p)$
in the approximation that higher-order terms provide only a
regulator $(1/R)$ for the pole at $p=\sqrt{6t_0/7s}$ in 
$p^2/V(P)$:
\eqn\resu{g_{ab}(\xi -\xi ') \simeq \delta_{ab}\ {1\over {7 s}}
\ \sqrt{2\over {\pi \sqrt{6t_0 \over 7s} |\xi -\xi '|}}
\ {\rm sin} \left( \sqrt{{6t_0 \over 7s}} 
|\xi -\xi '| - {\pi\over 4}\right) \ .}
To check the correctness of this result it is sufficient to compute
backwards its Fourier transform by first noting that \resu \ represents
the asymptotic behaviour of the von Neumann function
$(1/7s) N_0 \left( \sqrt{6t_0 /7s} |\xi -\xi '| \right) $.
The two-dimensional Fourier transform \gr \ of this function,
\eqn\back{\int d^2 \xi \ {1\over 7s} 
\ N_0 \left( \sqrt{6t_0 \over 7s} |\xi | \right) \ {\rm e}^{-ip\xi }
={1\over 2} \ {1\over {{3\over 4} t_0-{7\over 8} s \ p^2}}\ ,\qquad
0 < p < \sqrt{6t_0 \over 7s}\ ,}
reproduces exactly the small-$p$ behaviour of the momentum-space
correlator \ex .

The correlation function \resu \ is {\it long-range} in the usual
sense that $\int d^2\xi \ g_{aa} (\xi )$ is infrared divergent.
Strictly speaking, the large infrared cutoff $R$ should
be incorporated in the correlation function \resu . This can be removed
to infinity under the integral \back \ for all $p$ but $p=0$. So, there is
no finite correlation length and $\int d^2\xi \ g_{aa} (\xi )$ diverges
like $\sqrt{R\Lambda }$, a situation analogous to the
``Kosterlitz-Thouless order" in the low-temperature phase of the $O(2)$
non-linear sigma model \zinn .

Given that the vectors $\partial _a \phi ^i$ describe how the surface is
growing in and out of the $\left( \xi ^0, \xi ^1 \right) $ plane, the
oscillatory behaviour of \resu \ indicates that the surface fluctuates
around the reference plane with a wavelength
\eqn\wave{\ell (\tau )= 2\pi \sqrt{ 7s(\tau )\over 6t_0(\tau )} =
\sqrt{ {7\pi ^2\over 3} {K_1(\tau )\over \tau K_0(\tau ) }} \ .}
The scale of the amplitude, instead, is set by the parameter
\eqn\amp{a (z, \tau ) = {\sqrt{\ell (\tau )}\over 7 \pi s(z,\tau )} \propto
{1\over z^2} \ {\tau ^{3\over 4} \over {K_1(\tau )^{3\over 4} K_0(\tau )^{1
\over 4}}}\ .}
Given the behaviour
\eqn\beha{\eqalign{\tau &\ll 1 \ : \qquad
a \ll 1\ ,\cr
\tau &\gg 1\ : \qquad
a \gg 1\ ,\cr }}
combined with the fact that $a \propto 1/z^2$, we obtain a {\it smooth surface},
with waves of small amplitude, for small $\tau $ and large $z$: this 
corresponds to the perturbative domain derived above.
 
In order to confirm further this result we estimate the Hausdorff dimension
of the surface as a function of the parameters $z$ and $e$.
To this end we follow \david \ and compute the ratio $h\left( z, e, D
\right) $ between the expectation value of the 
(squared) distance $D^2_E$ of two points on the surface in embedding space
and its projection $D^2$ on the reference plane
$\left( \xi ^0, \xi ^1 \right)$. 

The (squared) distance in embedding space is the sum of the projection
$D^2$ on the reference plane and the contribution from normal
fluctuations:
\eqn\des{\eqalign{D^2_E &= D^2 + D^2_{\rm perp}\ ,\cr
D^2 &= |\xi -\xi '|^2\ ,\cr
D^2_{\rm perp} &= \sum_i \ \langle |\phi ^i (\xi )-\phi ^i (\xi ') |^2
\rangle \ ,\cr }}  
so that the ratio $h$ can be written as
\eqn\rah{h \left( z, e, D\right) = 1+ {D^2_{\rm perp} \left(
z, e, D\right) \over D^2}\ .}
The expectation value of the normal fluctuations can be easily computed
from \flu \ and \fouflu \ thereby obtaining
\eqn\finh{h\left(z, e, D\right) = 1+ {1\over \pi D^2}\ \int _0^1 dp\ \left(
1-J_0\left( pD\right) \right) \ {p\over V(p)}\ .}

In fig. 3 we plot the function $h\left( z, e, D\right) $ as a function
of $D$ for $z=10$ and $1/e = 0.22$ and $1/e = 0.5$. 

\vskip 30 pt                        
                         
\epsfbox{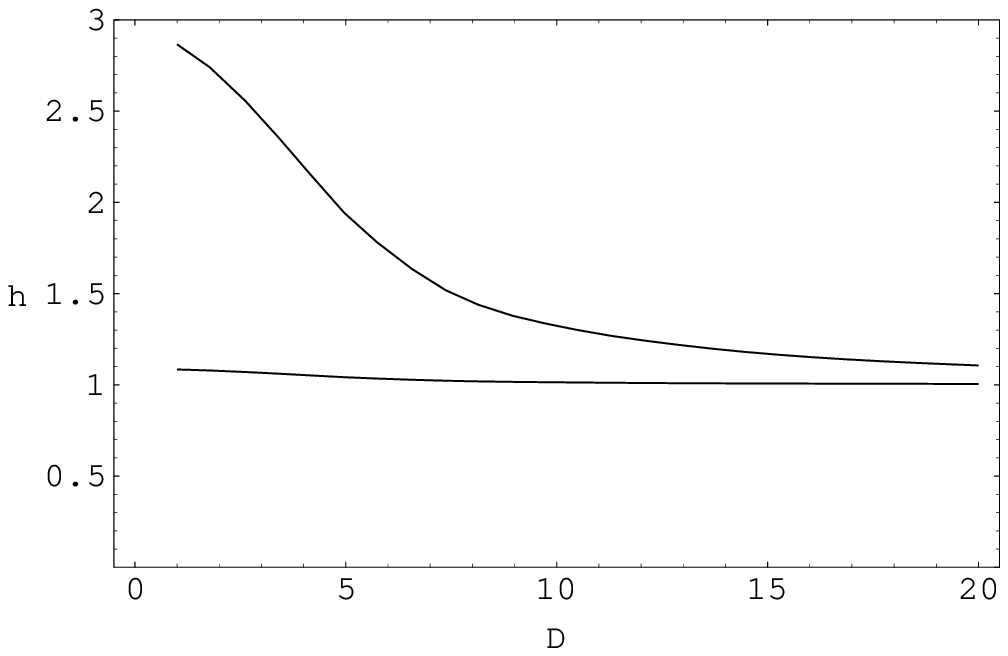}

\vskip 3pt
{\centerline {Fig. 3: the function $h$ for $1/e=0.22$ (lower curve)
and $1/e=0.5$ (upper curve).}}
\vskip 30pt

\noindent
The value $1/e=0.22$ 
corresponds to the boundary in fig.1, on which the perturbation parameter
$r(10, 1/(0.22))= 0.2$. We see that, at this value of $1/e$, the distance 
in embedding space still scales with the projected distance on the reference
plane, indicating that the Hausdorff dimension of the surface is indeed two.
At even lower values of $1/e$, corresponding to the interior of the
perturbative region the curve $h\left( z, e, D\right) $ is virtually
indistinguishable from the constant 1. For purpose of illustration we have
also plotted the function $h$ for $z=10$ and $1/e=0.5$, to investigate what
happens if $1/e$ is increased towards the non-perturbative region. The now
strong decrease of $h$ as a function of $D$ shows that the (squared) distance
in embedding space begins to scale slower than $D^2$ starting on small to
intermediate scales. This indicates that on these scales the surface is
loosing its intrinsic dimensionality two, thereby acquiring a higher
Hausdorff dimension. Note that the same picture applies to the other values
of $z$: for $e$ large enough we have scaling and Hausdorff dimension two;
when $e$ is lowered the surface begins to crease on small scales.
We conclude thus that, in the perturbative regime, confining strings
describe indeed smooth surfaces with intrinsic dimension two.

\newsec{Including a $\theta $-term}
When the underlying gauge theory contains a $\theta $-term
\eqn\gaueteta{S_{\rm gauge}^{\theta } = \int d^4x \ i{\theta \over 64 \pi ^2}
\ F_{\mu \nu }\epsilon_{\mu \nu \alpha \beta}F_{\alpha \beta}\ ,}
(in Euclidean space) the Kalb-Ramond action \akrf \ is modified as
follows \dqt :
\eqn\mod{S\left( B_{\mu \nu } \right) \to S\left( B_{\mu \nu } \right)
+ \int d^4x \ i{\theta \over 64\pi^2} \ B_{\mu \nu }\epsilon_{\mu \nu \alpha
\beta}B_{\alpha \beta}\ .}
Note that, for {\it compact} U(1) gauge fields, a $\theta $-term 
is not irrelevant since it produces non-trivial effects, notably it
assigns an electric charge $q=e\theta /2\pi$ to elementary magnetic
monopoles \wi , and these are responsible for the confining string.

This new term in the Kalb-Ramond action \akrf \ has two consequences
for the non-local formulation of the confining string action \nlf .
First of all, there is an additional term,
\eqn\modnonlo{S_{CS} \to S_{CS}+ i{z^2 \Lambda ^2 e^2 \theta \over
64\pi ^2} \ \int d^2{\sigma} \int d^2
{\sigma }' \ \epsilon_{\mu \nu \alpha \beta}
\ X_{\mu \nu}({\sigma}) \ Y\left( {\bf x}
({\sigma }) - {\bf x}({\sigma}') \right) 
\ X_{\alpha \beta}({\sigma }') \ ,}   
and, secondly, the mass \mass \ characterizing the Yukawa Green's function
$Y$ is modified to
\eqn\mteta{\eqalign{{m_{\theta }\over \Lambda} &= \tau _{\theta}
= {ez\over 4\pi }
\ \sqrt{\left( {4\pi \over e^2} \right) ^2 + t^2}\ ,\cr
t &\equiv {\theta \over 2\pi}\ .\cr }}
However, it is easy to convince oneself that the additional term in the
action does not contribute at all to the saddle-point expansion
\saddle \ to second order in the transverse fluctuations. Therefore,
the only consequence of the inclusion of a $\theta $-term is the
modification \mteta \ of the parameter $\tau $.

When $e\ll 1$, the new parameter $t$ is negligible, we regain the original
expression $\tau _{\theta}\simeq z/e$ and we end up in 
the non-perturbative regime.
When $e\gg 1$, $t$ plays a crucial role and we obtain $\tau _{\theta} \simeq
etz/4\pi $. This is the same expression as before with $e\to 4\pi /et$:
since now $e\gg 1$, however, we end up again in the non-perturbative regime.
We conclude that, in presence of a $\theta $-term, the smooth string can
exist only for {\it intermediate values} of $e$. It is worth noting, that
these intermediate values of $e$, for which $\tau $ is sufficiently small,
constitute exactly the non-perturbative regime of the gauge theory
with $\theta $-term, where dyons can condense \cara . Again, we recover the
duality between gauge fields and confining strings.

Note also that the restriction to intermediate values of $e$ prevents again
removing the cutoff and obtaining the Nambu-Goto string with self-intersection
term. This is entirely due to the one-loop corrections, which were not
included in \dqt .

\newsec{Large-$D$ analysis}
As we have pointed out in section 2, the two essential characteristics of the 
confining string action are that it represents a non-local interaction with
negative stiffness between surface elements, as best seen in the
formulation \ist . The range of the interaction is determined by the
parameter $\tau = z/e$ whereas the overall scale of the interaction 
depends only on $z$. 

The action \ist , however, has not the best form for the type of analysis
we have in mind.
Therefore, we shall consider a better-suited action which still incorporates
all essential aspects of confining strings. In dimensionless units this is:
\eqn\newa{S = \int d^2{\xi } \sqrt{g}\ \ g^{ab}
{\cal D}_a r_{\mu } \ W\left( z, e , \left( 
{{\cal D}\over \Lambda } \right) ^2
\right) \ {\cal D}_b r_{\mu } \ ,}
where ${\cal D}_a$ denote covariant derivatives along the surface.
In this formulation the non-local interaction is written in terms of the
tangent vectors $\partial _a r_{\mu }$ to the surface, instead of the tangent
tensor $t_{\mu \nu }$, as in \ist . While the physics of \newa \ is
essentially equivalent to \ist , we do not know which interaction $W$ in
\newa \ corresponds exactly to $G$ in \ist. An exact translation is possible,
however, if we are interested only in the first two terms of $W$:
\eqn\extra{W \left( z, e, 
\left( {{\cal D}\over \Lambda} \right) ^2 \right)
= {t_0\over 2} + s \left( {{\cal D}\over \Lambda } \right) ^2 + 
v \left( {{\cal D}\over \Lambda } \right) ^4 +\dots \ ,}
with $t_0$ and $s$ given in \expa and $(v/s)\ll 1$, $(v/t_0)\ll 1$ for
$\tau \gg 1$.
The formulation \newa \ is the one used by Kleinert and Chervyakov \kb \ in
their computation of the finite temperature free energy. 

At this point we use standard large-$D$ techniques along the lines of
\refs{\larged ,\david} .
We first introduce a (dimensionless) 
Lagrange multiplier matrix $\lambda ^{ab}$ 
to enforce the constraint $g_{ab}=\partial _ar_{\mu }\partial_br_{\mu }$,
\eqn\lamu{S \to S+ \int d^2\xi \sqrt{g} \ \ \lambda
^{ab} \left( \partial _a r_{\mu } \partial _br_{\mu } - g_{ab} \right) \ .}
We then parametrize the world-sheet in the Gauss map as
\eqn\gauss{r_{\mu } (\xi ) = \left( \xi _0, \xi _1, \phi ^i (\xi )
\right) \ ,\qquad \qquad i=2, \dots , D\ ,}
where $-\Lambda \beta /2\le \xi_0 \le \Lambda \beta /2$, 
$-\Lambda R/2 \le \xi ^1 \le \Lambda R/2$ and 
$\phi ^i(\xi )$ describe the ($D$-2) 
transverse fluctuations. 
With the usual isotropy Ansatz
\eqn\iso{g_{ab}=\rho \ \delta_{ab} \ ,\qquad \qquad \lambda ^{ab} =
\lambda \ g^{ab} \ ,}
for the metric and the Lagrange multiplier of infinite systems 
($\beta ,R \to \infty $) at the saddle
point we obtain
\eqn\gmapac{S =2\int d^2\xi \ \left( {t_0\over 2} +\lambda
(1-\rho ) \right) + \int d^2\xi  \ \partial_a\phi ^i
\left( \lambda + W\left( z, e , \left( {{\cal D}\over \Lambda }\right) ^2
\right) \right) \ \partial_a \phi ^i\ .}
Integrating over the transverse fluctuations, in the infinite
area limit, we get the effective action
\eqn\newact{S^{\rm eff} = 2\Lambda ^2 A_{\rm ext} \ \left( {t_0\over 2} +\lambda
(1-\rho ) \right) + \Lambda ^2 A_{\rm ext} {{D-2}\over 8\pi^2 }\rho
\int d^2p\ {\rm ln}
\left\{ p^2 \left( \lambda + W
\left( z, e, p^2\right) \right) \right\} \ ,}
where $A_{\rm ext}=\beta R$ is the extrinsic, physical area. For large
$D$, the fluctuations of $\lambda $ and $\rho $ are suppressed and these
variables take their classical values, determined by the two saddle-point
equations
\eqn\sapoi{\eqalign{\lambda &= {{D-2}\over {8\pi }}\ \int_0^1dp\ p\ {\rm ln}
\left\{ p^2 \left( \lambda + W 
\left( z, e, p^2\right) \right) \right\} \ ,\cr 
{{\rho-1}\over \rho} &= {{D-2}\over{8 \pi }}
\ \int _0^1 dp\ p\ 
{1\over {\lambda + W \left( z, e, p^2\right) }}\ ,\cr }}
where we have introduced again the ultraviolet regularization $p<1$.
Inserting the first saddle-point equation into \newact \ we get 
\eqn\efac{S^{\rm eff}= \Lambda ^2 \ \left( t_0+2\lambda \right)
\ A_{\rm ext} \ ,} 
from where 
we read off the renormalized string tension
\eqn\tension{\eqalign{T &= \Lambda ^2 \ t\ ,\cr
t &\equiv t_0+2\lambda \ .\cr }}

The physics of confining strings in the large-$D$ limit is determined thus
by the two saddle-point equations \sapoi . The first of these equations
requires the vanishing of the ``saddle-function" 
\eqn\safu{f(z, e, \lambda) \equiv \lambda - {{D-2}\over {8\pi }}
\ \int_0^1dp\ p\ {\rm ln}
\left\{ p^2 \left( \lambda + W 
\left( z, e, p^2\right) \right) \right\} \ ,}
and determines the Lagrange multiplier $\lambda $ as the solution of a
transcendental equation. The second saddle-point equation
determines then the metric, once the Lagrange multiplier has been found, and
can be written simply as
\eqn\lame{\rho = {1\over f'(z, e, \lambda )}\ ,}
where a prime denotes the derivative with respect to $\lambda $.

We start our analysis of the saddle-point equations by examining the weak
coupling case $e\ll z$ for fixed $z=0(1)$. In this case we have $\tau \gg 1$
and we can expand the potential $V$ in \newa \ as in \extra . Keeping only
the first, dominant term gives
\eqn\largeta{f(z, e, \lambda) = \lambda - {{D-2}\over {16 \pi }}
\left\{ {\rm ln} \left( \lambda + {t_0\over 2} \right) -1 \right\} \ .}
This function has a global minimum at
\eqn\glomin{\lambda^* = -{t_0\over 2} + {{D-2}\over {16 \pi }}\ ,}
at which the function takes the value
\eqn\vaglomin{f\left( z, e, \lambda ^* \right) = -{t_0\over 2} +
{{D-2}\over 16 \pi } \left( 2-{\rm ln} {{D-2}\over 16 \pi } \right) \ .}
For $\tau $ sufficiently large this expression is positive for $D=4$.
This shows, independently of the details of the interaction $W$, that for
sufficiently weak coupling $e\ll 1$ (and fixed $z$) there is no solution
to the saddle-point equations. The dominant large-$D$ approximation
admits a smooth confining string only for large coupling $e$, in full
agreement with the results of the perturbative analysis of section 3. 

Things become harder in the case $e\gg 1$. In order to shed light
on the complete quantum phase structure of confining strings
we shall resort to a toy model, by choosing the
specific interaction
\eqn\klemodel{W\left( z, e , \left( {{\cal D}\over \Lambda }\right) ^2
\right) = \bar W\left( z, e , \left( {{\cal D}\over \Lambda }\right) ^2
\right) = 
{z^2\over {\tau ^2 - \left( {{\cal D}\over \Lambda }\right) ^2 }}\ ,}
which is essentially the model of Kleinert and Chervyakov \kb .
As in \ist \ and \sposs\ we have two mass scales in addition to the
ultraviolet cutoff $\Lambda $. The mass $z\Lambda $ determines the
overall scale of the interaction between surface elements, whereas the
mass $z\Lambda /e$ determines the range of this interaction.
In this case we get
\eqn\gapkle{\eqalign{f(z, e, \lambda ) = 
\lambda &- {{D-2}\over 16 \pi}\left\{ -1
-\tau ^2 \ {\rm ln}\left( 1+{1\over\tau ^2}\right) - {{\tau ^2\lambda +z^2}
\over \lambda } \ {\rm ln}\left( {{\tau ^2\lambda +z^2 }\over {1+\tau ^2}}
\right) \right\} \cr 
&- {{D-2}\over 16 \pi}\left\{
{\tau ^2\lambda +z^2+\lambda \over \lambda }
\ {\rm ln} \left( \lambda +{z^2 \over {1+\tau ^2}} \right) \right\} \ .\cr }}
This function has the following limiting values:
\eqn\prope{\eqalign{&\lim _{\lambda \to \infty} f(z, e, \lambda ) 
=\infty \ ,\cr
&\lim _{\lambda \to \lambda _{\rm min}} = -z^2 -{{D-2}\over {16 \pi }} \left(
-1+{\rm ln} \ z^2 \right) + O\left( \tau ^2 \ {\rm ln} \ \tau ^2 \right) \ ,
\qquad \tau \ll 1\ ,\cr
&\lambda _{\rm min} = {-z^2 \over {1+\tau^2}} \ .\cr }}
For $z$ sufficiently large we have $\lim _{\lambda \to \lambda _{\rm min}}
<0$ and there exists at least one solution to the saddle-point equation
$f(z, e, \lambda )=0$. 

The derivative of the saddle-function $f$,
\eqn\dsf{f'(z, e, \lambda ) = 1-{{D-2}\over {16 \pi }}\left\{ {1\over \lambda}
-{z^2 \over \lambda ^2} \left( {\rm ln}\left( 1+{1\over \tau ^2} \right) +
{\rm ln} {{\lambda +{z^2 \over {1+\tau ^2}}} \over 
{\lambda +{z^2 \over \tau ^2}}}
\right) \right\} \ ,}
determines the metric element $\rho $ via \lame . Given that 
\eqn\proder{\eqalign{&\lim _{\lambda \to \infty} f'(z, e,\lambda )= 1\ ,\cr
&\lim_{\lambda \to \lambda_{\rm min}} f'(z, e, \lambda) = -\infty \ ,\cr }}
the saddle-function $f$ must have an odd number of extrema.
Our numerical analysis shows that it has 
exactly one minimum: when this minimum lies
above zero, the saddle-point equations have no solutions. When the minimum
lies below zero the saddle-point equation $f(z, e, \lambda)=0$ has two
solutions. Only the largest of these two solutions, however, is physical since
at the smallest one we have $f'(z, e, \lambda )=1/\rho <0$. So, in this case
we have exactly one physical solution of the saddle-point equations.

In fig. 4 we plot the critical line in parameter space below which there exists
one solution to the saddle-point equations for $D=4$. 

\vskip 30 pt                        
                         
\epsfbox{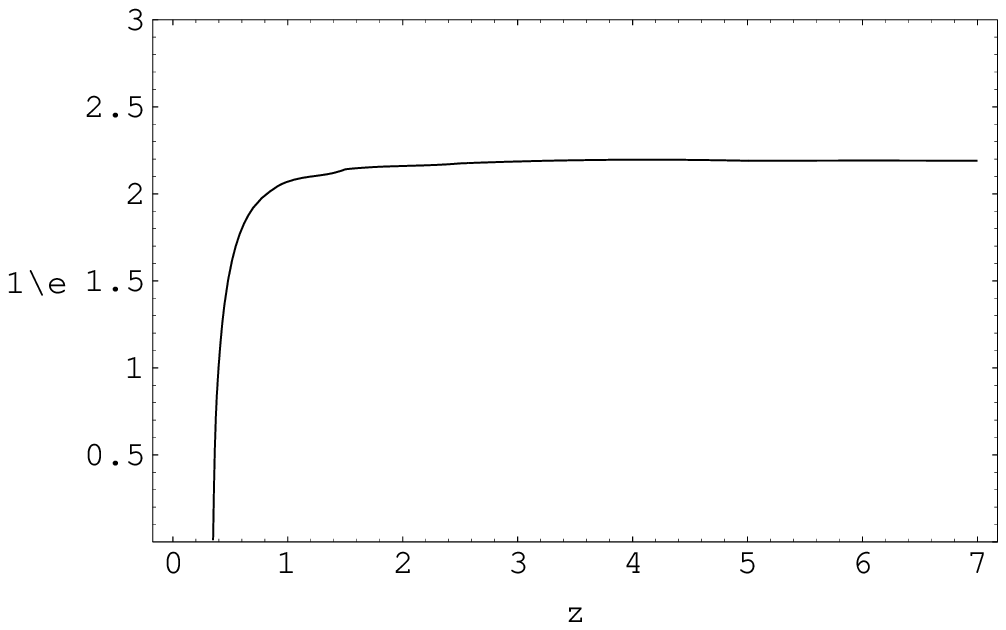}

\vskip 3pt
{\centerline {Fig. 4: the critical line in parameter space; smooth
strings exist below this line.}}
\vskip 30pt

\noindent
As expected, the region bounded by the critical line matches 
essentially the perturbative
region found with the one-loop calculation of section 3. Note, however, that,
in our toy model, there is no phase transition for large $z$ and fixed
(large) $e$. In this region the critical line $[1/e] \ (z) $ becomes a constant. 

We now 
compute the correlation function
\eqn\ladcorr{g_{ab}(\xi -\xi ') = \langle \partial_a \phi ^i (\xi )
\ \partial_b \phi ^i (\xi ') \rangle 
=\delta _{ab} 
\ {1\over (2\pi )^2}\ \int d^2p \ {1\over 
2\left( \lambda +\bar W\left( z, e, p^2\right) \right) }
\ \ {\rm e}^{i \sqrt{\rho } p (\xi -\xi ')} \ ,}
where $\bar W\left( z, e, p^2\right) $ is the Fourier transform of the
interaction in \newa . 
We start by rewriting this correlation function as
\eqn\nladcorr{g_{ab} 
=\delta _{ab} 
\ {1\over (2\pi )^2}\ \int d^2p \ \left[ {1\over 2\lambda} + 
{1\over {\delta \left( \bar t - 2\bar s \delta p^2 \right) }} \right]
\ \ {\rm e}^{i \sqrt{\rho } p (\xi -\xi ')} \ ,}
where 
\eqn\barqua{\eqalign{\bar t &\equiv 2\left( 
\lambda + {z^2\over \tau ^2} \right) \ ,\cr
\bar s &\equiv {z^2\over \tau ^4}\ ,\cr }}
are the (dimensionless) tension and stiffness, respectively and we
have introduced the new parameter $\delta \equiv |\lambda | \tau ^2/z^2$.
Also, we have used the fact that the saddle-point solution for $\lambda $
is negative. The first, constant term in \nladcorr \ corresponds to a
$\delta $-function contribution which we shall drop from now on. The 
second term, instead, can be treated with a computation completely
analogous to the one leading to \resu : 
\eqn\larlar{g_{ab}(\xi -\xi ') \simeq \delta_{ab}\ {1\over {8 \delta ^2\bar s}}
\ \sqrt{2\over {\pi \sqrt{{\bar t \rho} \over 2\delta \bar s} |\xi -\xi '|}}
\ {\rm sin} \left( \sqrt{{{\bar t \rho} \over 2\delta \bar s}} 
|\xi -\xi '| - {\pi\over 4}\right) \ ,}
where, as before, this form is valid up to a large infrared cutoff
$R$ such that $1/R$ regulates the pole in \nladcorr .
Again we find {\it long-range correlations} for the normal components of
tangent vectors to the world-sheet, indicating a smooth surface.

Let us now examine what happens when the critical line of fig. 4 is
approached from below.
In this case the minimum value of the saddle-function 
$f(z, e, \lambda )$ approaches zero. 
As a consequence, the solution of the saddle-point equation
$f(z, e, \lambda)=0$ coincides with the value $\lambda ^*$
where the function $f$ takes the minimum value. At this value
$\lambda ^*$, however, we have also 
$f' \left( z, e, \lambda ^* \right) =0$, so that,
due to \lame \ the metric element $\rho $ diverges. We 
conclude that 
\eqn\phtr{\lim_{(z, e) \to (z, e)_{\rm cr}} \rho \ \ = \infty \ .}
This means that, approaching the critical line, the ratio
$A_{\rm int}/A_{\rm ext}$ of the intrinsic to the extrinsic ($\beta R$)
area of the surface diverges. 
Moreover, since $\rho \to \infty$, the correlations
$\langle \partial_a \phi ^i (\xi ) \ \partial_b \phi ^i (\xi ') \rangle $ 
vanish for all $\xi \ne \xi '$.

We don't know the exact nature of the {\it phase transition} occuring on
the critical line depicted in fig. 4. The diverging intrinsic distances and
the vanishing correlators, however, indicate that surely the surface looses
its intrinsic dimensionality two and becomes at best a fractal object.
Note, however, that it is not a crumpling transition: in a crumpled phase
we should still have a solution to the saddle-point equations, even if
the correlations \ladcorr \ are short-range. It is not clear to us whether
a generic solution of the saddle-point equations, not restricted by the
ansatz \iso , exists beyond the critical line. Such a solution would
correspond to a fractal phase (possibly a branched-polymer phase \dfj )
of the string. The other possibility is that no solutions exist at all
beyond the critical line in which case all kind of strings would simply
be suppressed.  

\newsec{Conclusions}

We have shown that confining strings have smooth world-sheets with
long-range correlations in a perturbative region characterized by
strong gauge coupling $e$ and large monopole fugacity $z$. Decreasing
the coupling $e$ (at fixed fugacity $z$) or decreasing fugacity (at fixed
coupling $e$) a phase transition takes place at which the world-sheets
become at best fractal objects. Together with the facts that the confining
string theory describes a compact $U(1)$ gauge theory at strong coupling
\dqt \ and that its finite temperature free energy matches the large-$N$ 
QCD result both in temperature dependence and sign \kb , our results
make confining strings indeed very good candidates to describe 
the strong coupling phase of gauge theories.

\listrefs
\end